\documentclass[twocolumn,aps,showpacs,prb]{revtex4-1}

\usepackage{graphicx}
\usepackage{epstopdf}
\usepackage{amsmath}
\usepackage{color}
\usepackage{multirow}

\begin{document}

\title{Hydrogenation induced magnetic and electronic transitions in monolayer electride Gd$_2$C: A first-principles study}

\author{Duo Xu}
\author{Jian-Feng Zhang}
\author{Zhong-Yi Lu}\email{zlu@ruc.edu.cn}
\author{Kai Liu}\email{kliu@ruc.edu.cn}

\affiliation{Department of Physics and Beijing Key Laboratory of Opto-electronic Functional Materials $\&$ Micro-nano Devices, Renmin University of China, Beijing 100872, China}

\date{\today}

\begin{abstract}
The recently synthesized two-dimensional electride Gd$_2$C was proposed to be a ferromagnetic metal that possesses multiple pairs of Weyl points and may display a large anomalous Hall conductivity [Liu \textit{et al.}, Phys. Rev. Lett. \textbf{125}, 187203 (2020)]. In view of its layered structure, here we carry out first-principles studies on the magnetic and electronic properties of Gd$_2$C in the ultrathin monolayer limit. We find that monolayer Gd$_2$C remains ferromagnetic like the bulk form and the hydrogenation can effectively tune its magnetism and electronic structure. With one-sided coverage of hydrogen atoms, monolayer Gd$_2$C becomes a half-metal with one spin channel around the Fermi level. For two-sided hydrogenation, monolayer Gd$_2$C transforms to an antiferromagnetic insulator with a band gap of 0.8 eV. Our studies show that monolayer electride Gd$_2$C can perform multiple magnetic and electronic transitions with different levels of hydrogenation and may be also adopted to construct a planar heterojunction with selective area adsorption of hydrogen atoms, which has promising applications in future electronic and spintronic devices.
\end{abstract}


\maketitle

\section{INTRODUCTION}

The electride materials are deemed to possess excess electrons, so called free electron gas, serving as anions~\cite{review1,review2}. These excess electrons locate on the lattice voids but not bind to certain nuclei~\cite{review1,review2}. According to the dimensionality of the free electron gas in the electrides, they can be classified into zero-dimensional (0D), one-dimensional (1D), two-dimensional (2D), and three-dimensional (3D) types~\cite{maprx}. The formation of an interstitial free electron gas is distinguished from the conventional covalent, ionic, or metallic bonds and can bring many interesting properties such as the low work function~\cite{lwf1,lwf2}, high electronic mobility~\cite{hem}, high electron concentration~\cite{hec}, high density of active sites~\cite{as}, thermionic electron emission at low temperature~\cite{lt}, to name a few. On the other hand, some electrides can also demonstrate rich quantum phenomena such as metal-semiconductor transition~\cite{mst}, superconductivity~\cite{sup}, magnetism~\cite{y2c}, nontrivial topological properties~\cite{review1,review2}, etc. Because of their particular quantum phases and potential applications in widespread fields~\cite{wf1,wf2}, the electride materials deserve more experimental and theoretical explorations.

A recent study~\cite{1stnc} reported the successful synthesis of a ferromagnetic 2D electride Gd$_2$C, which has a Curie temperature of $\sim$350 K. %, being much higher than that ($\sim$7 K) of Y$_2$C~\cite{y2c}.
The ferromagnetism of Gd$_2$C mainly comes from Gd-4$f$ electrons and the measured local moment on Gd is about 7.26 $\mu_\text{B}$~\cite{Gdmom}. In comparison, the electronic states around the Fermi level are mainly contributed by Gd-5$d$ electrons and the free electron gas~\cite{1stnc}, which are also spin polarized. A previous theoretical study predicted that the ferromagnetic Gd$_2$C can host multiple pairs of Weyl points around the Fermi level and exhibit a giant anomalous Hall conductivity (AHC)~\cite{gd2cprl}, which may have potential applications in future spintronic devices. In addition, a recent computational study suggested that an external electric field could alter the number of anionic electrons at the surface of few-layer Gd$_2$C~\cite{gd2cnew}. With the reduced dimensionality, whether the physical properties of Gd$_2$C can be retained in the ultrathin limit and can be tuned via surface modification still need investigation.

In this work, by using first-principles calculations, we have studied the magnetic and electronic properties of the electride material Gd$_2$C in its monolayer limit. Our calculations show that monolayer Gd$_2$C is still a ferromagnetic metal, resembling its bulk form. We have also investigated the influence of hydrogenation on monolayer Gd$_2$C to examine the possible magnetic transitions like hydrogenated graphene~\cite{Hgraphene1,Hgraphene2}, hydrogenated monolayer Ca$_2$N~\cite{qiujpcc}, and chlorin-intercalated bulk Gd$_2$C~\cite{1stnc}. We find that the one-sided hydrogenation can completely suppress one spin channel around the Fermi level and drives Gd$_2$C to a ferromagnetic half-metal, while the two-sided hydrogenation can transform Gd$_2$C to an antiferromagnetic insulator. The monolayer electride Gd$_2$C can thus perform multiple magnetic and electronic transitions at different levels of hydrogenation.

\section{Method}

To investigate the electronic and magnetic properties of monolayer Gd$_2$C with hydrogenation, we carried out the density functional theory (DFT)\cite{dft1,dft2} calculations by using the Vienna Ab-initio Simulation Package (VASP)\cite{vasp1,vasp2}. The generalized gradient approximation (GGA) of Perdew-Burke-Ernzerhof (PBE) \cite{PBE} type was adopted for the exchange-correlation functional. A kinetic energy cutoff of 520 eV was used for the plane-wave basis. A $\bf {k}$-point mesh of 16$\times$16$\times$1 was adopted for the Brillouin zone (BZ) sampling in the structural optimization calculations, and a find grid of 60$\times$60$\times$1 in the self-consistent calculations. The Gaussian smearing method with a width of 0.02 eV was used for the Fermi surface broadening. To describe the strong correlation effect among Gd-4$f$ electrons, the GGA + U formalism~\cite{ldau} was adopted with an effective Hubbard U of 6.0 eV, the same as a previous theoretical study~\cite{gd2cprl}.
In phonon calculations, the real-space force constants were calculated within the density functional perturbation theory (DFPT)\cite{dfpt} as implemented in VASP\cite{vasp1,vasp2} and the phonon dispersion was then calculated with Phonopy\cite{phonopy}. The AHC of monolayer Gd$_2$C was studied within the Berry curvature scheme~\cite{VanderbiltPRB}, which was implemented in Wannier90~\cite{mlwf} and WannierTools~\cite{wt} packages: $\sigma_{xy}=-\frac{e^2}{\hbar}\int_{\text {BZ}}\frac{dk_x dk_y}{(2\pi)^2}\Omega_{xy}(k)$. Here $\Omega_{xy}(k)=\sum_n{f_n(k)\Omega_{n,xy}(k)}$ is the total Berry curvature~\cite{VanderbiltPRB}, where $f_n(k)$ is the occupation factor and $n$ is the index of energy band. The spin-orbit coupling (SOC) effect was included in the AHC calculations.

\section{Results and Discussion}

\begin{table}[b]
\caption{The relative energies (in units of meV/f.u.) of typical magnetic configurations (Fig. \ref{fig5} in the Appendix) with respect to their own FM states for different Gd$_2$C monolayers: $\Delta E = E - E$(FM). Here FM and AFM are abbreviations of ferromagnetic and antiferromagnetic, respectively. The FM-FM means all Gd atoms are in ferromagnetic intralayer and interlayer couplings. The Stripe and Zigzag stand for the stripe AFM and zigzag AFM intralayer couplings in the same Gd layers. The abbreviations "semi-hydro." and "full-hydro." represent the semi-hydrogenation with one-sided H coverage and the full-hydrogenation with two-sided H coverage, respectively. }
\begin{center}
\begin{tabular*}{8cm}{@{\extracolsep{\fill}} cccc}
\hline\hline
$\Delta E$ & pristine & semi-hydro. & full-hydro. \\
\hline
FM-FM & 0.0 & 0.0 & 0.0 \\
FM-AFM & 164.5 & 81.1 & -28.2 \\
Stripe-FM & 194.5 & 97.2 & -28.2 \\
Stripe-AFM & 169.8 & 101.7 & -16.5 \\
Zigzag-FM & 155.8 & 83.3 & -25.8 \\
Zigzag-AFM & 158.5 & 94.3 & -23.1 \\
%Bistripe-FM & 106.9 & 44.9 & -18.2 \\

\hline\hline
\end{tabular*}
\end{center}
\end{table}

\begin{figure}[t]
\includegraphics[angle=0,scale=0.6]{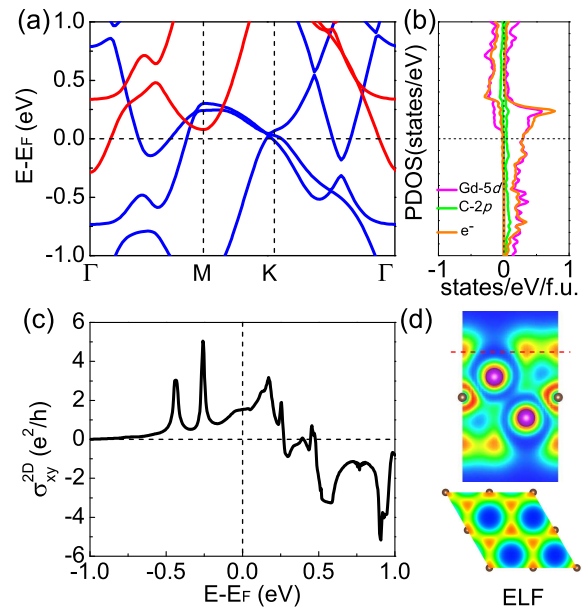}
\caption{(Color online) (a) Electronic band structure of pristine Gd$_2$C monolayer. The red and blue lines represent the spin-up and spin-down channels, respectively. (b) Partial density of states (PDOS) of pristine Gd$_2$C monolayer. The pink, green, and orange lines label the states from Gd-5$d$, C-2$p$, and anionic electrons, respectively. (c) Calculated 2D anomalous Hall conductivity (AHC). (d) Electron localization function (ELF) distributions in the (110) and (001) planes. The latter is at a position above the surface labeled by the red dashed line. {}}
\label{fig1}
\end{figure}

The calculated cleavage energy of Gd$_2$C is 1.21 J/m$^2$, which is slightly larger than that of Ca$_2$N (1.09 J/m$^2$)\cite{Ca2N}. The comparable cleavage energies between these two electrides indicate that Gd$_2$C can be also reduced to the monolayer thickness via the liquid exfoliation method as Ca$_2$N\cite{druffel}. According to the calculated phonon dispersion [Fig. \ref{fig4}(a)], the pristine Gd$_2$C monolayer has no imaginary frequency through the entire BZ, indicating its dynamical stability. Since the Gd atom contains seven $4f$ electrons, we studied several typical magnetic configurations of the Gd$_2$C monolayer (Fig. \ref{fig5} in the Appendix), whose relative energies are listed in Table I. As can be seen, the ferromagnetic (FM) state is energetically most favorable for the pristine Gd$_2$C monolayer, being the same as its bulk crystal~\cite{gd2cprl}. The total magnetic moment of the unit cell is 15.94 $\mu_B$, mainly distributing on the Gd atoms. Moreover, the calculated energy differences between the FM and nonmagnetic states for monolayer and bulk Gd$_2$C are -19.08 and -21.94 eV per formula unit (f.u.) respectively. %, mainly contributed by Gd atoms.
%In consideration of the Curie temperature $T_\text{c}$ of 350 K for bulk Gd$_2$C, we deduce that the $T_\text{c}$ of monolayer Gd$_2$C may also exceed \textcolor{red}{???} K. In comparison, the Curie temperature $T_\text{c}$ is only 45 K in monolayer CrI$_3$~\cite{CrI3}.
These results suggest that Gd$_2$C can retain stable and ferromagnetic in the monolayer limit~\cite{ChaePRB}.

Figure \ref{fig1}(a) shows the band structure of pristine Gd$_2$C monolayer in the FM ground state, where both the spin-up (red lines) and spin-down (blue lines) channels cross the Fermi level, indicating its metallic characters. From the partial density of states [Fig. \ref{fig1}(b)], both the Gd-5$d$ orbitals and the electron gas floating on the surface [Fig. \ref{fig1}(d)] have large contributions around the Fermi level. Interestingly, from the top view of the electron localization function (ELF) distribution in Fig. \ref{fig1}(d), the electron gas forms an approximate hexagonal pattern, while the crystal structure is still in C$_3$ symmetry. This hexagonal pattern contains two lattice sites: one is on top of C atom, the other is on top of the lower-Gd atom. In comparison, the free electron gas in bulk Gd$_2$C condenses in the interlayer region and forms a triangular pattern in the ELF map~\cite{1stnc}. As bulk Gd$_2$C was predicted to own a large intrinsic AHC with $\sigma_{xy}=399~\Omega^{-1}$cm$^{-1}$ at $E_\text{F}$~\cite{ahc}, we also investigated the intrinsic 2D AHC $\sigma_{xy}^{\text {2D}}$ of pristine Gd$_2$C monolayer. From Fig. \ref{fig1}(c), the calculated $\sigma_{xy}^{\text {2D}}$ at the Fermi level is about 1.8 $e^2/h$ (= $0.698$$\times$$10^{-4}$ $\Omega^{-1}$), which is larger than the ones of some traditional 2D ferromagnetic materials such as Mn$_3$P~\cite{mn3p}.

In order to regulate the magnetic and electronic properties, we have applied the lattice strain on the pristine Gd$_2$C monolayer. It turns out that when a 5\% biaxial tensile strain is applied, the total magnetic moment of the unit cell increases slightly to 16.02 $\mu_B$ without introducing remarkable changes in the electronic band structure. This suggests that the magnetism of the pristine Gd$_2$C monolayer are not so sensitive to the lattice strain. Next, we try to adjust the physical properties of monolayer Gd$_2$C by other means.

\begin{figure}[t]
\includegraphics[angle=0,scale=0.6]{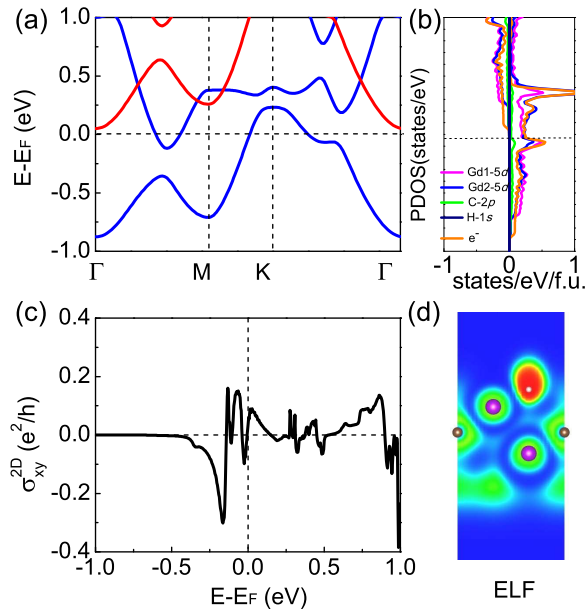}
\caption{(Color online) (a) Electronic band structure of the semi-hydrogenated Gd$_2$C monolayer. The red and blue lines represent the spin-up and spin-down channels, respectively. (b) PDOS of the semi-hydrogenated Gd$_2$C monolayer. The pink, blue, green, and orange lines represent the states from the 5$d$ orbitals of upper and lower Gd atoms, C-2$p$ orbitals, and anionic electrons, respectively. (c) Calculated 2D AHC. (d) The ELF distribution in the (110) plane vertical to the Gd$_2$C monolayer. }
\label{fig2}
\end{figure}

It is well known that hydrogenation is a common way to tune the electronic and magnetic properties of layered materials. For example, the hydrogenation can drive the MoO$_3$ nanoribbon from a semiconductor to a metal~\cite{moo3}, induce the structural and magnetic transitions in Ca$_2$N monolayer~\cite{qiujpcc}, transform arsenene into a Dirac material~\cite{arsenenes}, as well as induce high-temperature superconducting phases in iron-based superconductors~\cite{febase}. In our study, because there is free electron gas locating on both surfaces of the Gd$_2$C monolayer, hydrogenation will be an effective method to modulate the properties.

We first investigated the coverage of hydrogen atoms on one side of Gd$_2$C monolayer, namely the semi-hydrogenation case. Multiple adsorption sites of H atom were examined and the optimal site is on top of the lower-Gd atom [Fig. \ref{fig2}(d)]. The adsorption energy at this site is 124.9 meV/H larger than at the other site (on top of C) of the approximate hexagonal lattice in Fig. \ref{fig1}(d). The dynamical stability of the semi-hydrogenated Gd$_2$C monolayer is also confirmed by the phonon calculation, in which no imaginary frequency is found in the whole BZ [Fig. \ref{fig4}(b)]. From the relative energies in Table I, we find that the optimal magnetic configuration of semi-hydrogenated Gd$_2$C monolayer is still the FM state. The calculated band structure of the FM state is exhibited in Fig. \ref{fig2}(a). Notably, only one spin channel (blue lines) crosses the Fermi level, forming a ferromagnetic half-metal. By analyzing the projected DOS [Fig. \ref{fig2}(b)], we find that the two Gd atoms are not equivalent any more due to the asymmetrical adsorption of H atoms on one side. %Meanwhile, both Gd atoms and the electron gas are spin-polarized.
The calculated ELF distribution in Fig. \ref{fig2}(d) indicates that the free electron gas on one side of the Gd$_2$C monolayer transfers to the adsorbed H atoms and the electron gas on the other surface is retained. In order to explore the electronic transport properties, we also calculated the intrinsic AHC $\sigma_{xy}^{\text {2D}}$ as shown in Fig. \ref{fig2}(c). Compared with the pristine case [Fig. \ref{fig1}(c)], the AHC in the semi-hydrogenated Gd$_2$C monolayer decreases obviously. This variation may originate from the reduced number of Weyl points in Fig. \ref{fig2}(a), which weakens the magnitudes of Berry curvature $\Omega_{xy}(k)$ and AHC.

\begin{figure}[tb]
\includegraphics[angle=0,scale=0.5]{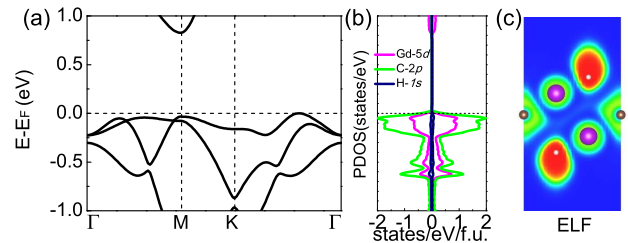}
\caption{(Color online) (a) Electronic band structure and (b) PDOS of the fully-hydrogenated Gd$_2$C monolayer in the FM-AFM state. The pink and green colors label the states from Gd-5$d$ and C-2$p$ orbitals. The signs represent different spin channels. (c) The ELF distribution in the (110) plane vertical to the Gd$_2$C monolayer. }
\label{fig3}
\end{figure}

\begin{figure*}[tb]
\includegraphics[angle=0,scale=1.0]{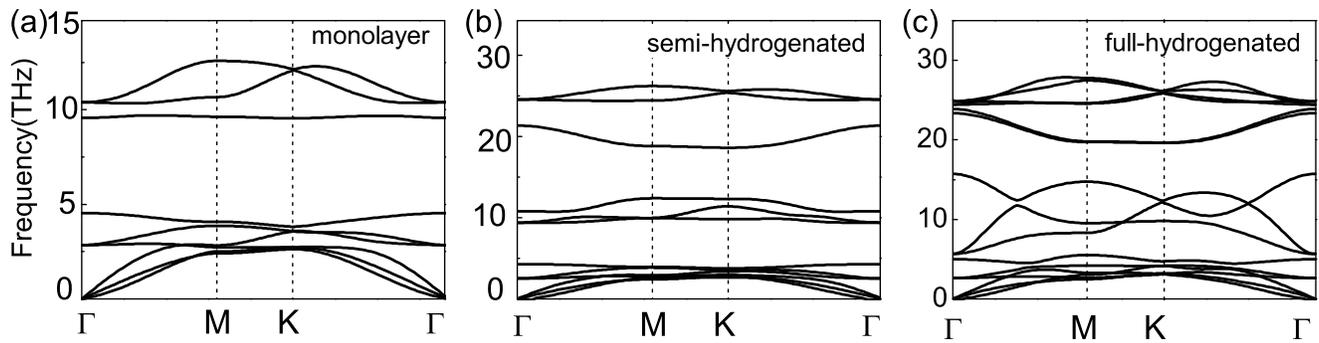}
\caption{(Color online) Calculated phonon dispersions of the (a) pristine, (b) semi-hydrogenated, and (c) fully-hydrogenated Gd$_2$C monolayers.}
\label{fig4}
\end{figure*}

When both sides of monolayer Gd$_2$C are covered with hydrogen atoms, the system becomes fully hydrogenated. According to the calculated phonon dispersion [Fig. \ref{fig4}(c)], it is still dynamically stable. By comparing the total energies of different magnetic configurations (Table I), we find that the magnetic ground state of the fully-hydrogenated Gd$_2$C monolayer is either of the intra-layer ferromagnetic coupling and the inter-layer antiferromagnetic coupling [Fig. \ref{fig5}(b)] or of the intra-layer stripe AFM coupling and the inter-layer FM coupling [Fig. \ref{fig5}(c)]. The energy degeneracy of these two states originates from the equal number of AFM superexchange interactions between the Gd atoms in the upper and lower Gd layers via the bridging C atoms, namely one C atom forming three AFM Gd(upper)-C-Gd(lower) bonds with six nearest neighboring Gd atoms around it. The magnetism all comes from the inner Gd-4$f$ shell with a 7.09 $\mu_B$ local moment, resulting in a zero net magnetic moment for the whole cell. From the band structures of the AFM ground state shown in Figs. \ref{fig3}(a) and \ref{fig5}(a), the fully-hydrogenated Gd$_2$C monolayer is an insulator with a band gap of $\sim$0.8 eV, in sharp contrast to the pristine and semi-hydrogenated cases. This is because all the free electron gases on the surfaces of Gd$_2$C monolayer are now localized around the adsorbed H atoms, as indicated by the ELF distribution in Figs. \ref{fig3}(c) and \ref{fig5}(c). According to the PDOS [Figs. \ref{fig3}(b) and \ref{fig5}(b)], C-$2p$ orbitals dominate the valence-band maximum, while Gd-$5d$ orbitals determine the conduction-band minimum.
%In addition, it can be seen from Table 1 that, there is another magnetic structure with the same lowest energy in the fully hydrogenated monolayer Gd$_2$C system, namely the Stripe-FM. In this two structures, one C atom forms three antiferromagnetic Gd-C-Gd bonds with six nearest neighbour Gd atoms around it, and the band structure of  Stripe-FM also indicates that it is an antiferromagnetic insulator with a band gap of 0.81 eV.
%Thus the chemical formula of the system can be denoted as Gd$_2$CH$_2$, while the 5$d$ and 6$s$ electrons of the Gd atoms mostly transfer to the C or H atoms.
The above results indicate that the Gd$_2$C monolayer transforms into an antiferromagnetic insulator under the full hydrogenation.
%It is worth mentioning that, among the five antiferromagnetic structures listed in Table 1, the number of antiferromagnetic bonds formed by C atom and its nearest neighbor Gd is 3, 2, 1 and 0, with the corresponding energy differences with ferromagnetic states are -28.2, -25.8, -23.1 and -16.5 meV respectively. This indicates that the formation of antiferromagnetic Gd-C-Gd bonds is more energetically advantageous in the system of fully hydrogenated monolayer Gd$_2$C.

Finally, we studied the magnetic anisotropy of the Gd$_2$C monolayer without and with the hydrogenation. In the pristine Gd$_2$C monolayer, the easy magnetization axis is along the out-of-plane direction. With the destruction of central-inversion symmetry by the coverage of H atoms on one side, the easy magnetization axis of semi-hydrogenated Gd$_2$C monolayer turns to in-plane as a result of the Rashba-type spin-orbit coupling~\cite{rashba}. The values of magnetic anisotropy energy (MAE) are 0.663 and 0.236 meV/Gd in pristine Gd$_2$C monolayer and semi-hydrogenated Gd$_2$C monolayer respectively. After H atoms cover on both sides, the magnetic anisotropy energy of the system becomes negligible, which may originate from the reduced interaction between localized Gd-4$f$ electrons in the insulating state for fully-hydrogenated Gd$_2$C.

\section{Discussion and Summary}

The monolayer electride Gd$_2$C and its hydrogenated forms have many advantages. First, the pristine Gd$_2$C monolayer owns a stable FM state and may display a large AHC. Second, the semi-hydrogenated Gd$_2$C monolayer is an FM half-metal, which may be used as a spin filter to generate a 100\% pure spin current. Third, the fully-hydrogenated Gd$_2$C monolayer is an AFM insulator (I) that can realize the high switch ratio. Last but not least, with a selective area adsorption of hydrogen atoms, the Gd$_2$C monolayer can be processed to planar heterojunctions with the FM/I/FM modules. The diverse functionalities of the hydrogenated Gd$_2$C monolayer may thus have promising applications in future electronic and spintronic devices.

%Different from other 2D magnetic electride the distinct changes of electronic properties of Gd$_2$C material with the change of atomic chemical environment caused by adsorption H atoms which from single side H to double side H underwent a strange phase transition from ferromagnetic metal to ferromagnetic semi-metal and then to antiferromagnetic insulator. The changes in electron structure brought about by this unusual structural phase transition may have more interesting potential applications and enriched the research field of two-dimensional materials. Besides, other than the anomalous Hall effect of other layered two-dimensional materials measured experimentally~\cite{experahc}, the anomalous Hall conductance of our system is spontaneously caused by symmetry break due to the adsorption of H atoms instead of the condition of applied electric field. This enlightenment as a unique intrinsic material in electronic devices will play a unique advantage.

In summary, we have performed the first-principles electronic structure calculations on the ferromagnetic electride material Gd$_2$C in the monolayer limit. Our calculations indicate that the pristine Gd$_2$C monolayer is still a ferromagnetic metal, the same as its bulk form. Interestingly, we find that the coverage of hydrogen atoms on one side can suppress one of the spin channels and drive the Gd$_2$C monolayer into a ferromagnetic half-metal. After the full hydrogenation with two-sided coverage of H atoms, the Gd$_2$C monolayer will further transfer to an antiferromagnetic insulator. We also calculate the anomalous Hall conductivity for the pristine and semi-hydrogenated Gd$_2$C monolayers, where the former shows a large value of 1.8 $e^2/h$, superior to traditional 2D ferromagnetic materials. Further calculations indicate that the pristine Gd$_2$C monolayer has an out-of-plane easy axis of magnetization and the semi-hydrogenated one owns an in-plane easy axis. In contrast, the magnetic anisotropy is insignificant for the full hydrogenation case. The versatile magnetic and electronic properties induced by the hydrogenation of the monolayer electride Gd$_2$C may thus be applied to future miniaturized electronic and spintronic devices.

\begin{acknowledgments}

This work was supported by the National Key R\&D Program of China (Grants No. 2017YFA0302903 and No. 2019YFA0308603), the National Natural Science Foundation of China (Grants No. 11774424, No. 11934020, and No. 12174443), the Beijing Natural Science Foundation (Grant No. Z200005), the Fundamental Research Funds for the Central Universities, and the Research Funds of Renmin University of China (Grant No. 19XNLG13). Computational resources were provided by the Physical Laboratory of High Performance Computing at Renmin University of China.

\end{acknowledgments}

\begin{appendix}
\section{}

\begin{figure}[thb]
\includegraphics[angle=0,scale=0.25]{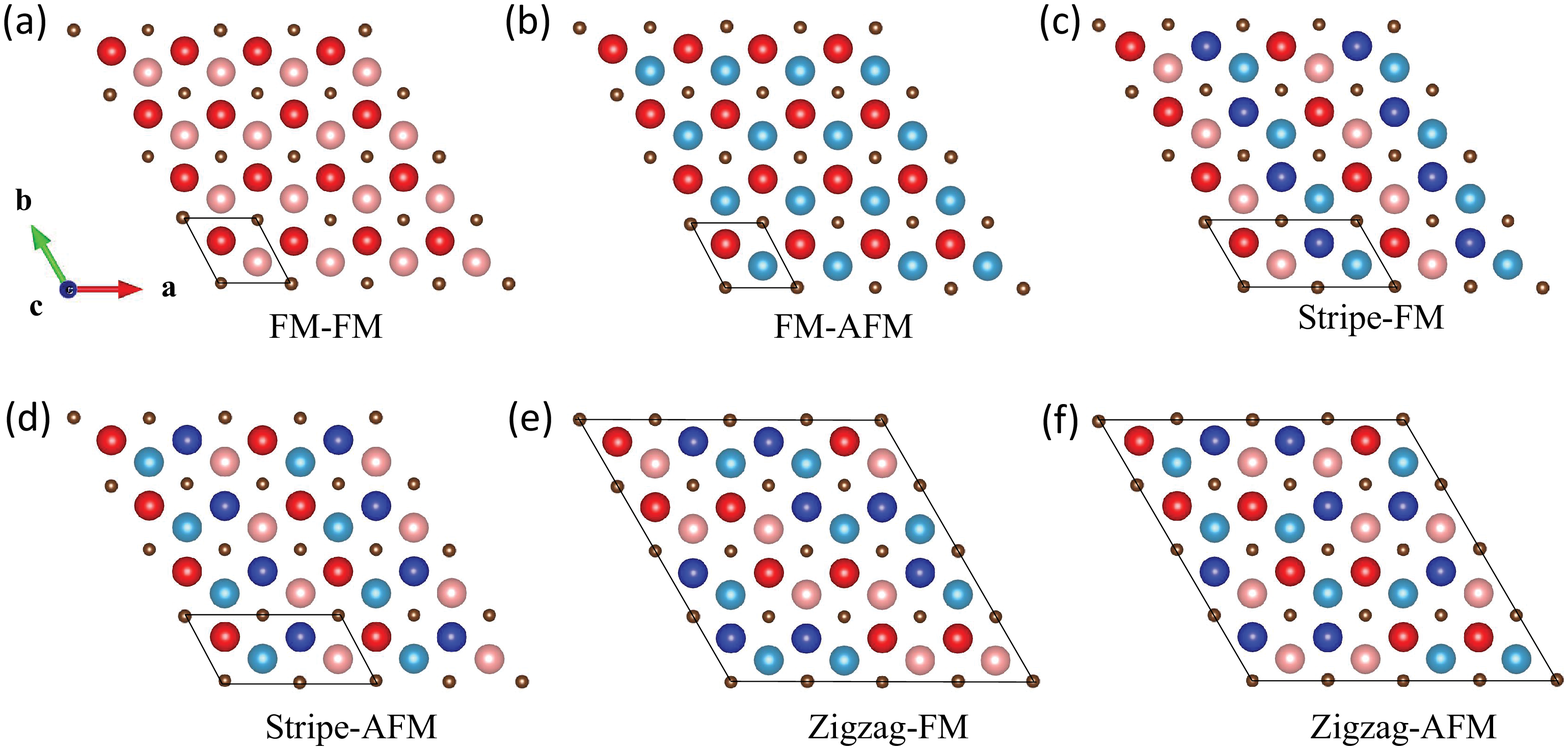}
\caption{Typical magnetic configurations for the Gd triangular lattices in monolayer Gd$_2$C. The red (blue) and light-red (light-blue) balls represent the spin-up (spin-down) Gd atoms in the upper and lower Gd layers, respectively. The brown balls represent the C atoms in the middle layer. The solid parallelograms label the unit cells of these magnetic structures. }
\label{fig5}
\end{figure}

Figure \ref{fig5} displays the typical magnetic configurations for the Gd triangular lattices in the Gd$_2$C monolayer, whose relative energies are listed in Table I.

\begin{figure}[thb]
\includegraphics[angle=0,scale=0.5]{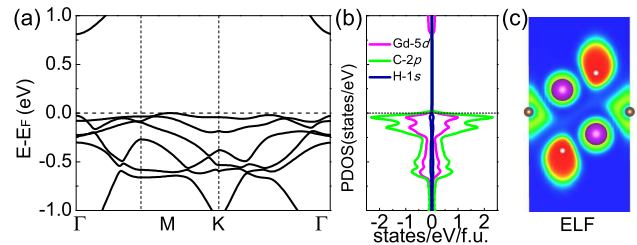}
\caption{(Color online) (a) Electronic band structure and (b) PDOS of the fully-hydrogenated Gd$_2$C monolayer in the Stripe-FM state (Table I). The pink and green colors label the states from Gd-5$d$ and C-2$p$ orbitals. The signs represent different spin channels. (c) The ELF distribution in the (110) plane vertical to the Gd$_2$C monolayer. }
\label{fig6}
\end{figure}

Figure \ref{fig6} shows the electronic structure of the fully-hydrogenated Gd$_2$C monolayer in the Stripe-FM state (Table I), which is also an AFM insulator as the FM-AFM state (Fig. \ref{fig3}).

\end{appendix}

\end{document}